\documentclass[%
aip,
rsi,
 amsmath,amssymb,
reprint,%
]{revtex4-1}
\usepackage{graphicx}
\usepackage{dcolumn}
\usepackage{bm}
\usepackage[utf8]{inputenc}
\usepackage[T1]{fontenc}
\usepackage{mathptmx}
\usepackage{etoolbox}
\usepackage{comment}
\makeatletter
\providecommand*{\diff}%
		{\@ifnextchar^{\DIfF}{\DIfF^{}}}
\def\DIfF^#1{%
		\mathop{\mathrm{\mathstrut d}}%
				\nolimits^{#1}\gobblespace}
\def\gobblespace{%
				\futurelet\diffarg\opspace}
\def\opspace{%
		\let\DiffSpace\!%
		\ifx\diffarg(%
			\let\DiffSpace\relax
		\else
			\ifx\diffarg[%
				\let\DiffSpace\relax
			\else
				\ifx\diffrag\{%
					\let\DiffSpace\relax
				\fi\fi\fi\DiffSpace}
\def\@email#1#2{%
 \endgroup
 \patchcmd{\titleblock@produce}
  {\frontmatter@RRAPformat}
  {\frontmatter@RRAPformat{\produce@RRAP{*#1\href{mailto:#2}{#2}}}\frontmatter@RRAPformat}
  {}{}
}%
\makeatother
\begin{document}
\title{\textbf{Mass-selected Ion-molecule Cluster Beam Apparatus for Ultrafast Photofragmentation Studies}}



\author{Xiaojun Wang}
\thanks{Author to whom correspondence should be addressed: wang.xiaojun@huskers.unl.edu}
\affiliation{Department of Physics and Astronomy, University of Nebraska-Lincoln, Lincoln, NE 68588, USA.}

\author{Mahmudul Hasan}
\affiliation{Department of Physics and Astronomy, University of Nebraska-Lincoln, Lincoln, NE 68588, USA.}

\author{Lin Fan}
\affiliation{Department of Physics and Astronomy, University of Nebraska-Lincoln, Lincoln, NE 68588, USA.}

\author{Yibo Wang}
\affiliation{Department of Physics and Astronomy, University of Nebraska-Lincoln, Lincoln, NE 68588, USA.}

\author{Hui Li}
\affiliation{Department of Chemistry, Nebraska Center for Materials and Nanoscience, and Center for Integrated Biomolecular Communication, University of Nebraska-Lincoln, Lincoln, NE 68588, USA.}

\author{Daniel S. Slaughter} 
\thanks{Author to whom correspondence should be addressed: dsslaughter@lbl.gov}
\affiliation{Chemical Sciences Division, Lawrence Berkeley National Laboratory, 1 Cyclotron Rd, Berkeley, CA 94720, USA.}

\author{Martin Centurion}
\thanks{Author to whom correspondence should be addressed: martin.centurion@unl.edu}
\affiliation{Department of Physics and Astronomy, University of Nebraska-Lincoln, Lincoln, NE 68588, USA.}

\date{\today}

\begin{abstract}
We describe an apparatus to study the fragmentation of ion-molecule clusters triggered by laser excitation and transfer of an electron from the iodide to the neutral molecule. The apparatus comprises a source to generate ion-molecule clusters, a time-of-flight (TOF) spectrometer and a mass filter to select the desired anions, and a linear-plus-quadratic reflectron mass spectrometer to discriminate the fragment anions after the femtosecond laser excites the clusters. The fragment neutrals and anions are then captured by two channeltron detectors. The apparatus performance is tested by measuring the photofragments: I$^-$, CF$_3$I$^-$ and neutrals from photoexcitation of the ion-molecule cluster CF$_3$I$\cdot$I$^-$ using femtosecond UV laser pulses with a wavelength of 266 nm. The experimental results are compared with our ground state and excited state electronic structure calculations and the existed results and calculations, with particular attention to the dynamics of the photoexcitation and photodissociation.
\end{abstract}
\maketitle 
\section{Introduction}

Cluster research provides a way to use molecular clusters to model the microscopic behavior of chemical reactions masked out by the averaging effect of the bulk\cite{castleman_ionic_1986}. Clusters have been used as a means of preparing intermediate reactants like I$^-$$\cdot$CH$_3$I of the classic S$_N$2 reaction I$^-+$CH$_3$I in gas phase\cite{cyr_observation_1992}. Based on this unique role, clusters combined with photoelectron imaging have been used to explore charge transfer reactions\cite{cyr_observation_1992,cyr_photoelectron_1993,cyr_charge_1994,dessent_dipolebound_1995,dessent_observation_1995,dessent_vibrational_1996} and the effects of solvation on the energetics and dynamics of chemical reactions\cite{mabbs_photoelectron_2005}. As a successful photoelectron imaging method\cite{doi:10.1146/annurev.pc.46.100195.002003}, velocity mapping imaging (VMI) introduced by Eppink and Parker\cite{eppink_velocity_1997} has been widely used in chemical dynamics research of neutrals and anions\cite{stolow_femtosecond_2004,surber_probing_2003}. The detection of photofragments including fragment-ions and neutrals is equally important for understanding the dissociation dynamics of ion-molecule clusters after excitation triggered by pulsed laser\cite{dessent_vibrational_1996}.

As well as electron imaging, VMI can be used to detect fragment-ions, but it is typically limited in the range of accessible kinetic energies, e.g., below a few eV for high resolution imaging of a gas phase chemical reaction\cite{doi:10.1146/annurev.pc.46.100195.002003,mikosch_imaging_2008,wester_velocity_2013}. In cluster ion beams, however, parent ion-molecule clusters are accelerated to a few thousand eV to be mass-separated using a Wiley-Mclaren time of flight (TOF) spectrometer\cite{wiley_timeflight_1955}, then mass-selected and transported to photoexcitation region\cite{alexander_recombination_1988}. The resultant dissociating photofragment ions often have kinetic energies from a few hundred to a few thousand eV. A reflection mass spectrometer\cite{alexander_recombination_1988,oh_tandem_2004} is one standard method to discriminate the photofragment ions after photoexcitation of the parent ion-molecule cluster of a few thousand eV. 

In this article we describe an experimental instrument designed to investigate the excited-state dissociation dynamics of mass-selected ion-molecule clusters by mass-resolving and detecting photofragment-ions and neutrals. In our setup, we have a source to generate ion-molecule clusters, a TOF mass spectrometer and a mass filter to select the target ion-molecule cluster, and a linear-plus-quadratic reflectron\cite{oh_tandem_2004} (LPQR) to mass-resolve a broad range of anion fragments, detected in parallel after photoexcitation. Neutral fragments are detected on a separate detector in coincidence with anion fragments. We have carried out experiments on the CF$_3$I$\cdot$I$^-$ ion-molecule cluster photoexcited by 266 nm UV femtosecond laser pulses to demonstrate the instrument performance. We compare the experimental results with our electronic structure calculation and the prior work and discuss the dynamics of the photoexcited cluster.

\section{Apparatus} 

\begin{figure*}
\centering
\includegraphics[width=1.0\textwidth]{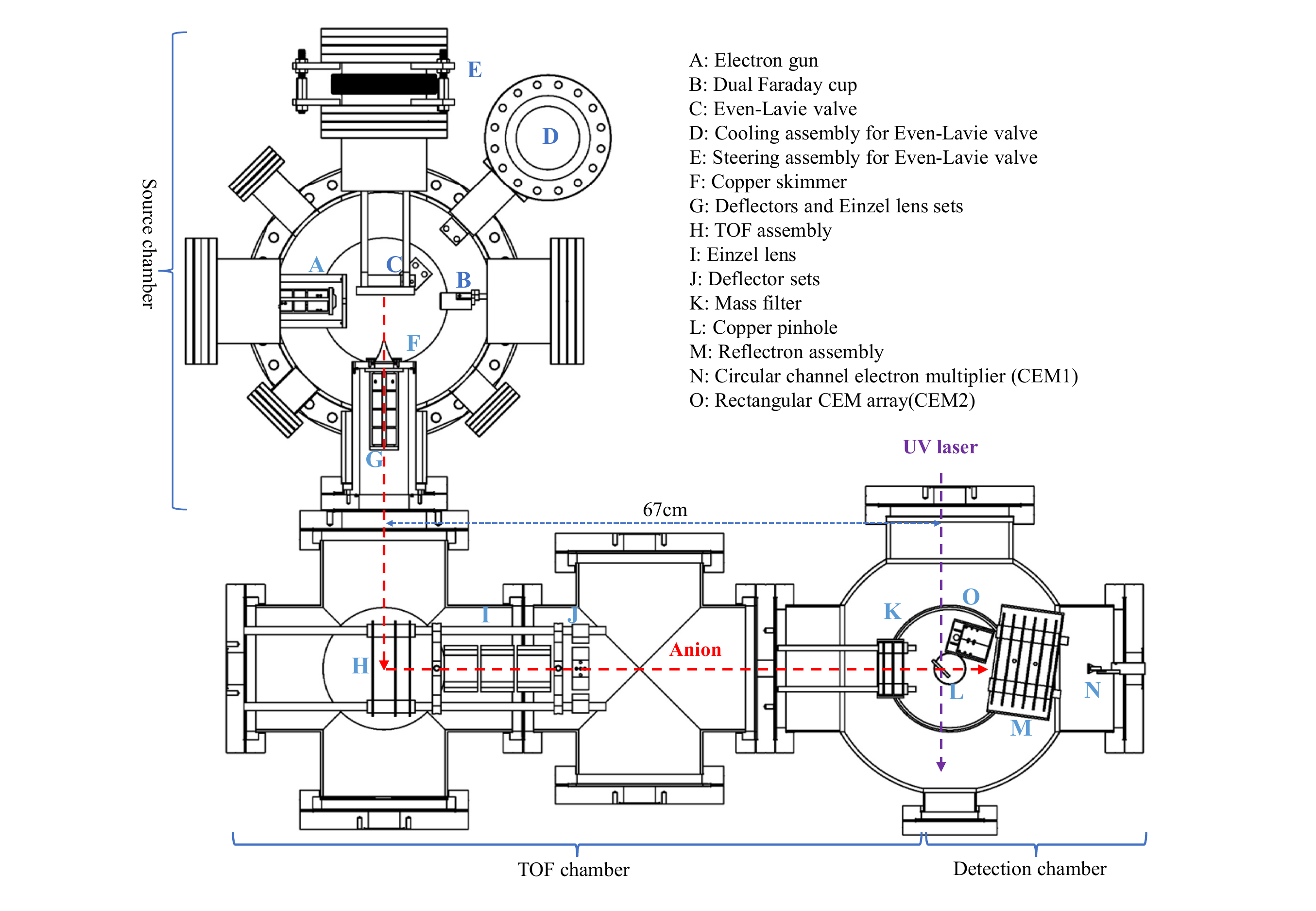}
\caption{Overview of the main components of the apparatus, consisting of the source chamber, TOF chamber and detection chamber. The gas manifold system, pump system and electronic control system are not shown here. }
\label{overview}
\end{figure*}

We have developed an apparatus to generate ion-molecule clusters by electron attachment in a gas jet, followed by TOF mass selection. The specified ion-molecule cluster is photoexcited by a laser pulse and the resulting neutral and ionic fragments are analyzed by mass. The apparatus consists of three modules (Fig.~\ref{overview}): a source chamber where a gas mixture is expanded into ultrahigh vacuum and irradiated with electrons to create anion clusters; a TOF chamber for acceleration of the anion clusters, and mass-selection to isolate a specific cluster; and a detection chamber for neutral fragment detection and mass-analysis of anion fragments. The source chamber houses an electron gun and a pulsed supersonic expansion gas jet for the generation of ion-molecule clusters. The TOF chamber contains a spectrometer and ion optics used to guide, focus, and separate the clusters into mass-specific pulses as well as a mass filter to select the target ion-molecule cluster. The detection chamber contains the interaction volume of the laser and the ion-molecule cluster, a reflectron TOF mass spectrometer, and detectors to capture neutral fragments and mass-resolved ions. A timing system was built to synchronize the laser and anion pulses. 

\subsection{Source chamber: ion-molecule cluster generation}

\begin{figure}
\centering
\includegraphics[width=0.48\textwidth]{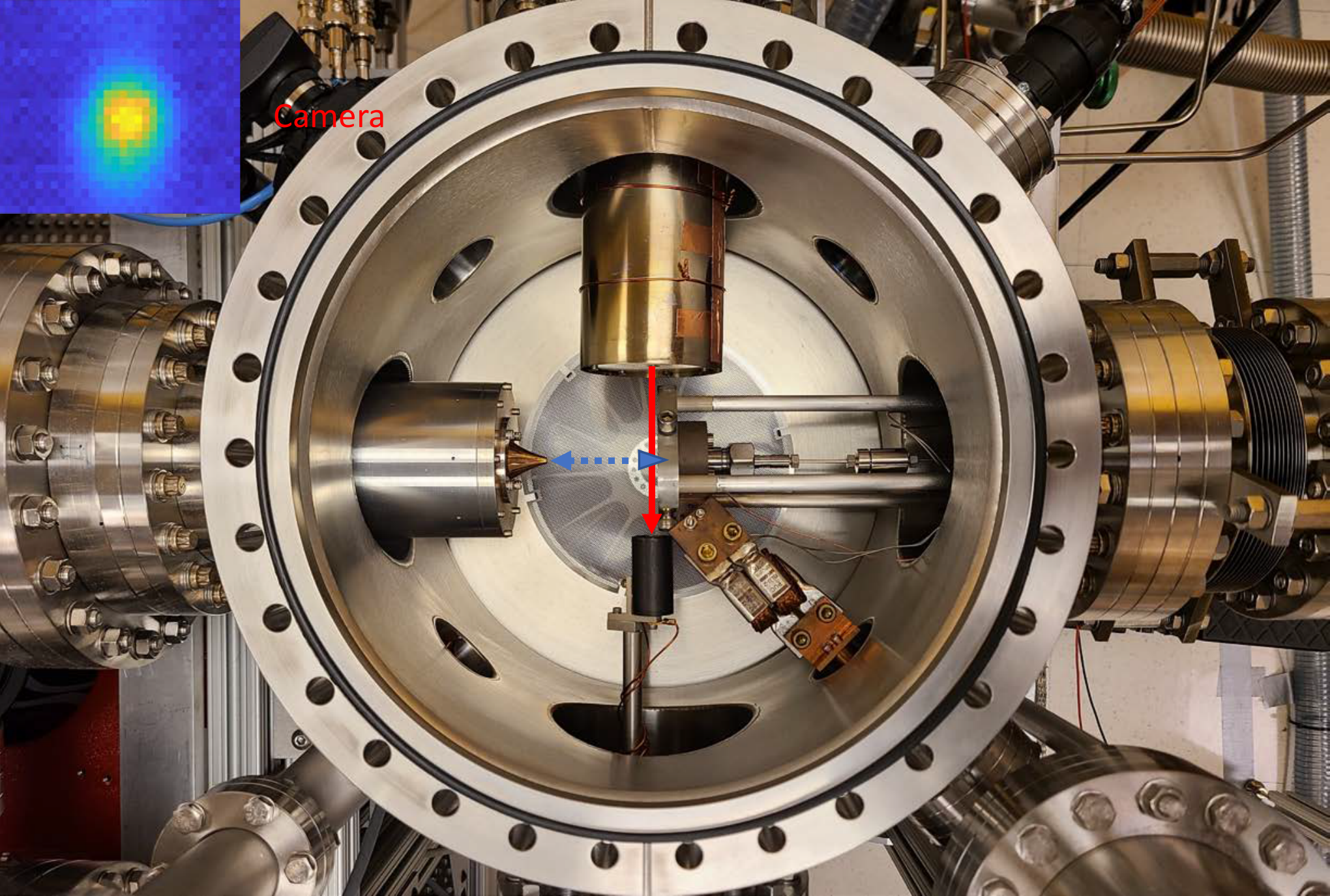}
\caption{Top view of the source chamber. Red solid line and blue triangle indicate electron beam and gas jet, respectively, and the dashed blue arrow represents the ion pulses generated by ionization. Insert shows a typical CMOS camera image of fluorescence emitted from the gas excited by the electron beam.}
\label{source}
\end{figure}

Ions and ion clusters are generated by irradiating a gas mixture containing the target molecule with an ionizing electron beam. Fig.~\ref{source} shows a photograph of the source chamber where a pulsed gas jet is crossed by the continuous electron beam in the center. The main components inside the source chamber are also displayed: home-made electron gun, dual Faraday cup, Even-Lavie pulsed valve\cite{even_even-lavie_2015}, cooling and steering assembly for the pulsed valve, copper skimmer and ion optics used to guide ions to next chamber, which are also shown in Fig.~\ref{overview}. Attached to the source chamber is a gas manifold system used to prepare the gas mixture, and the ultrahigh vacuum pump system used to pump down source chamber (not shown here). 

A homemade electron gun is employed to deliver a focused electron beam that ionizes the gas mixture near the exit of the pulsed valve nozzle. The electron gun uses the Pierce design where a tungsten filament is held at the same potential as a conical cathode\cite{pierce_rectilinear_1940}. An anode with a 1 mm aperture confines the size of the electron beam after thermionic emission from the filament. A three-element einzel lens, along with horizontal and vertical deflectors, is used to focus and guide the beam to the ionization region, at $\sim$1 mm from the exit of the pulsed valve. This design produces a continuous well-focused electron beam with a current of 400 {\textmu}A. The energy of the electron beam is set to 800 eV, which maximizes the ion yields within the constraints of the experimental geometry. The interface between the electron gun and the source chamber is a cylindrical aluminum shield with a 6 mm hole as the exit of electron beam, so that it is isolated from the source chamber. The electron gun is pumped by a dedicated turbopump to increase the operating lifetime of the tungsten filament. A sheet of {\textmu}-metal is wrapped around the cylindrical shield to decrease ambient magnetic fields inside the electron gun. A dual Faraday cup is used to monitor and stop the electron beam. It consists of two coaxial cups with diameters of 4 mm and 20 mm, insulated by Kapton film. We used the dual Faraday cup to estimate the size of electron beam based on the currents measured from the two cups. Additionally, we use a camera to image the optical emission from the ionization volume. When combined, these measurements provide an accurate measurement of the electron beam size.  

A gas mixture of 100 psi is delivered to a nozzle with an opening of 100 {\textmu}m by a gas manifold system. The gas manifold system consists of a reservoir of high-pressure gas mixture (1.3\% CF$_3$I and 98.7\% Ar), a 2-stage pressure regulator and a pulsed valve. The pressure regulator stabilizes and maintains the stagnation pressure of 100 psi behind the nozzle within 0.025\%. The pulsed valve is mounted on a stainless-steel bellow with tip-tilt and linear translation controls, such that the orientation of the pulsed valve, and the distance between the pulsed valve and the electron beam are adjustable. The pulsed valve is tightly wrapped with a copper belt that is connected to a cold finger by copper braids. The cold finger can be cooled by liquid nitrogen in an attached reservoir. A thermocouple attached to the valve is used to monitor the valve temperature.

After supersonic expansion through the pulsed valve nozzle, an intense pulsed gas jet is generated and crossed by the focused electron beam coming perpendicularly. The process involves electron-impact ionization\cite{mark_fundamental_1982} and low-energy electron attachment\cite{Illenberger1994}, and generates cations and anions including the ion-molecule cluster CF$_3$I$\cdot$I$^-$ and the iodide ion I$^-$. The ionization region is imaged onto a CMOS camera to monitor the overlap and the size of the electron and gas beams. We optimize the overlap of the gas jet and the e-beam by steering the electron beam with the electrostatic optics inside the electron gun and pulsed valve orientation with the bellow while monitoring the optical emission at the camera in real time. The size of the overlap region is calculated by fitting a two-dimensional Gaussian to the image of the plasma fluorescence (insert of Fig.~\ref{source}): typically, 2.43 mm$\times$3.23 mm at full width of half maximum (FWHM). 

The generated ions pass through a flared-conical (trumpet-like) copper skimmer with an entrance diameter of 3 mm which collimates the beam and allows for differential pumping of the source and TOF chambers. Planar horizontal and vertical  deflectors and an einzel lens are used to transport and focus the ions at the entrance to the TOF mass spectrometer. The typical flight time for the ion-molecule cluster CF$_3$I$\cdot$I$^-$ is 0.47 ms from the ionization region of the source chamber to the extraction region of the TOF assembly. The skimmer and ion optics are mounted on a cylindrical aluminum  tube which separates the source chamber from the TOF chamber. With the pulsed valve running at 200 Hz and 100 psi stagnation pressure, the pressures inside the source chamber and electron gun are $1-2\times10^{-5}$~Torr and $2-3\times10^{-6}$~Torr, respectively. 

\subsection{TOF chamber: separation and selection of ion-molecule cluster} 

The ions formed in the source chamber are transported through a 6~mm aperture, and focused at the entrance of the Wiley-McLaren style TOF mass spectrometer\cite{wiley_timeflight_1955} in the TOF chamber (H in Fig.~\ref{overview}). The TOF chamber accommodates the TOF mass spectrometer, a second einzel lens, a second set of planar deflectors, and a mass filter. The TOF mass spectrometer is used to extract and accelerate the ion pulses, dispersing ions of different mass in time. The TOF mass spectrometer can run in two modes, anions or cations, depending on the polarity of the potentials applied to the electrodes of the spectrometer. The einzel lens and deflectors focus the ion beam transversely and direct the beam to the laser interaction region in the detection chamber. The mass filter is deployed near the temporal focal position of the TOF mass spectrometer to select the ion-molecule cluster of interest, presently CF$_3$I$\cdot$I$^-$, while rejecting other ions.

The TOF mass spectrometer consists of three flat ring electrodes: repelling, extraction, and ground, which define an extraction region with a length of 25.4 mm and an acceleration region with a length of 15 mm. Each of the three electrodes is machined from 316 stainless steel, with 114 mm outer diameter and 3.175 mm thickness. The repelling electrode is a solid piece while the other two have an inner aperture with a diameter of 25 mm. The inner aperture is covered with 79~\% high-transparency stainless steel meshes to mitigate the lensing effect of the aperture\cite{liebl2008applied}. High voltage pulses with a duration of 4 {\textmu}s and voltages of -3000 V and -2535 V are applied to the repelling and the extraction electrodes, respectively. The TOF mass spectrometer accelerates the ions in a direction perpendicular to their incoming velocity. The HV pulse applied to the extraction electrode is delayed by 174 ns relative to that of the repelling electrode. This delay eliminates most of the energy spread introduced by the initial energy of ions and reduces the duration of the ion pulses from 400 ns to 80 ns (FWHM). A constant potential of -1525 V is applied to the middle element of the einzel lens to focus anion pulses transversely. In addition, a pair of parallel plate deflectors is used to steer the beam and compensate for the small initial transverse velocity component of the ions. Each anion species in the resulting anion beam is focused both transversely and longitudinally at the laser interaction point, 67 cm downstream from the center of the extraction region, with a time of arrival that depends on the mass. The operating pressure of the TOF chamber is typically $5\times10^{-7}$~Torr.

\begin{figure*}
\centering
\includegraphics[width=1.0\textwidth]{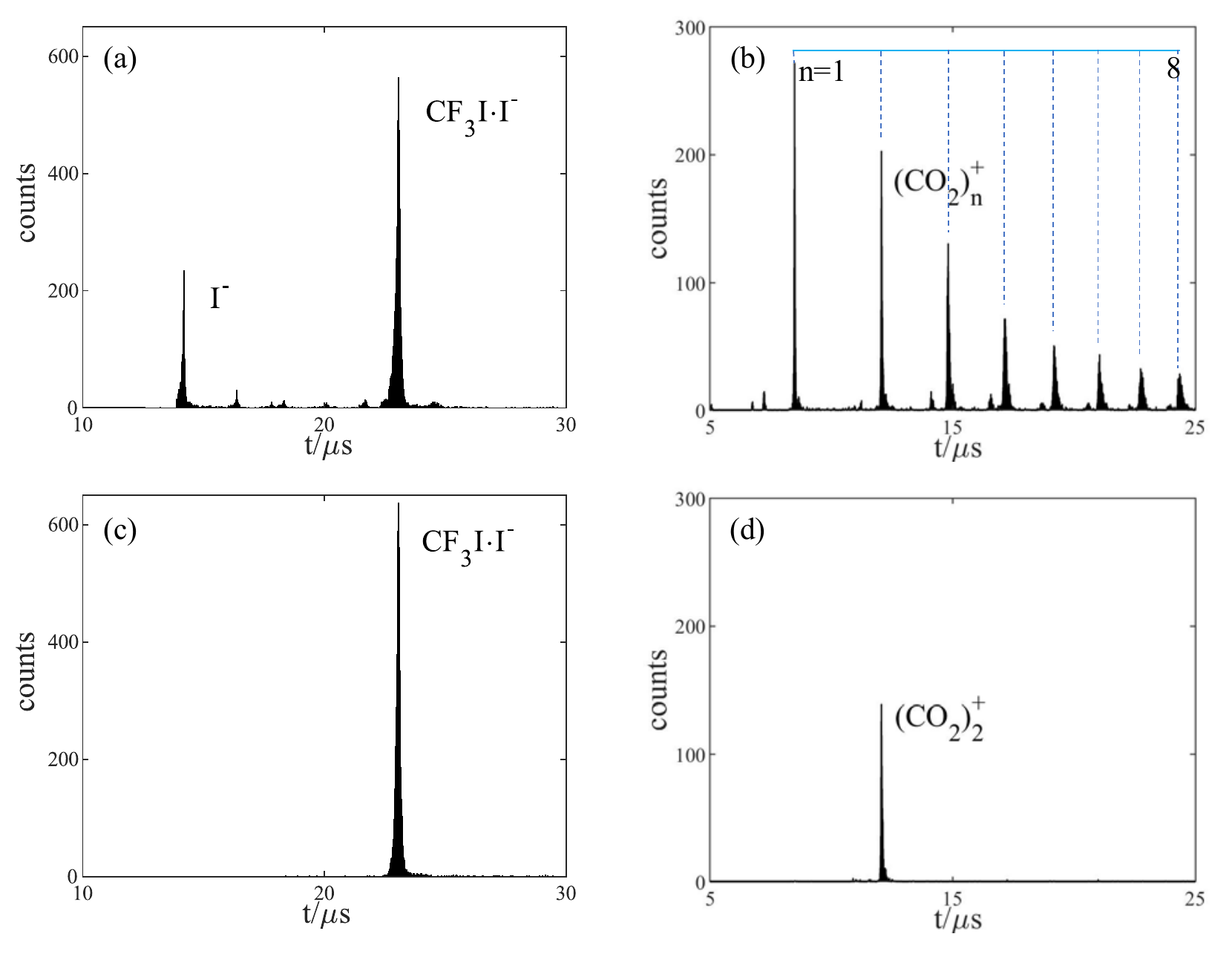}
\caption{TOF spectrum for (a) anions created from a gas mixture of Ar and CF$_3$I, without the mass filter, and (b) TOF spectrum of cations generated when CO$_2$ is used as source, without the mass filter. Blue dashed vertical lines indicate the (CO$_2$)$_n^+$ cation clusters from n=1 to n=8. (c) Same conditions as (a), but with the mass filter selecting the anion-molecule cluster CF$_3$I$\cdot$I$^-$. (d) Same conditions as (b), but with the mass filter selecting the cation cluster (CO$_2$)$_2^+$.}
\label{tof_mass_filter}
\end{figure*} 

The arrival time of the ion pulses is recorded using a channel electron multiplier (CEM) detector CEM1 placed at the end of the detection chamber (Fig.~\ref{overview}), which provides the TOF measurement. Fig.~\ref{tof_mass_filter}(a) shows a typical TOF spectrum of anions generated using a mixture of Ar and CF$_3$I. The most prominent peaks are identified as I$^-$ and CF$_3$I$\cdot$I$^-$, in addition to small traces of I$^-$$\cdot$(CH$_3$CN)$_n$, with n$>1$. Fig.~\ref{tof_mass_filter}(b) shows the TOF spectrum of the cations generated when CO$_2$ with 50 psi stagnation pressure is used as source, and the TOF mass spectrometer is in cation mode by reversing the polarity of the voltages. (CO$_2$)$_n^+$ cluster signals are recorded from n=1 to n=8 within the temporal range of the TOF spectrum, which is limited to 20 $\mu$s in this measurement. The mass resolution (m/$\Delta$m) of the TOF mass spectrometer is calculated to be 75 based on the spectrum of the CO$_2$ cation clusters shown in Fig.~\ref{tof_mass_filter}(b). The resolution here is more than sufficient to separate ion-molecule clusters as showed in Fig.~\ref{tof_mass_filter}(a) and Fig.~\ref{tof_mass_filter}(b).

A mass filter is employed to select the target ion-molecule cluster and reject all others before they reach the interaction region. The mass filter comprises three stainless steel ring electrodes with 60~mm outer diameter, 16~mm inner diameter and 1.6~mm thickness. A high-transparency conductive mesh is inserted to cover the aperture at the center of each electrode. A floating voltage of -3000 V is applied to the middle electrode to repel the ion clusters with two outer electrodes grounded. As the target ion-molecule cluster arrives at the entrance of mass filter, the floating voltage of -3000 V abruptly changes to 0 V within tens of ns and holds 0 V for 1 {\textmu}s while the target ion-molecule cluster passes through. The voltage changes back to -3000 V to repel the coming ion clusters as the target ion-molecule cluster exits the mass filter completely. Thus, only the target ion-molecule is permitted to enter the interaction region. A cylindrical aluminum  tube enclosing the mass filter electrodes is used to shield the detection chamber and CEM circuits from the pulsed electric fields inside the mass filter. The total length of mass filter must be smaller than the spatial separation between the selected ion and neighboring ions, or they will not be effectively separated. The mass filter is placed as close as possible to the interaction region to achieve the highest possible mass resolution. 

The HV pulse to the mass filter is carefully timed such that the TOF spectrum and the temporal shape of the ion pulses are not distorted. We have found a delay of 16.590 {\textmu}s to be optimal for selecting CF$_3$I$\cdot$I$^-$ clusters. Fig.~\ref{tof_mass_filter}(c) shows the selected target CF$_3$I$\cdot$I$^-$ in the TOF spectrum with the mass filter on, from which we estimate 90\% transmission of CF$_3$I$\cdot$I$^-$. The cation cluster selection of (CO$_2$)$_n^+$ is also shown in Fig.~\ref{tof_mass_filter}(d) using a time delay of 8.600 {\textmu}s to select (CO$_2$)$_2^+$. No counts above the noise floor are detected at the position of the rejected peaks, suggesting a rejection close to 100\%. 

\subsection{Detection chamber: photoexcitation of ion-molecule cluster}

\begin{figure*}
\centering
\includegraphics[width=0.96\textwidth]{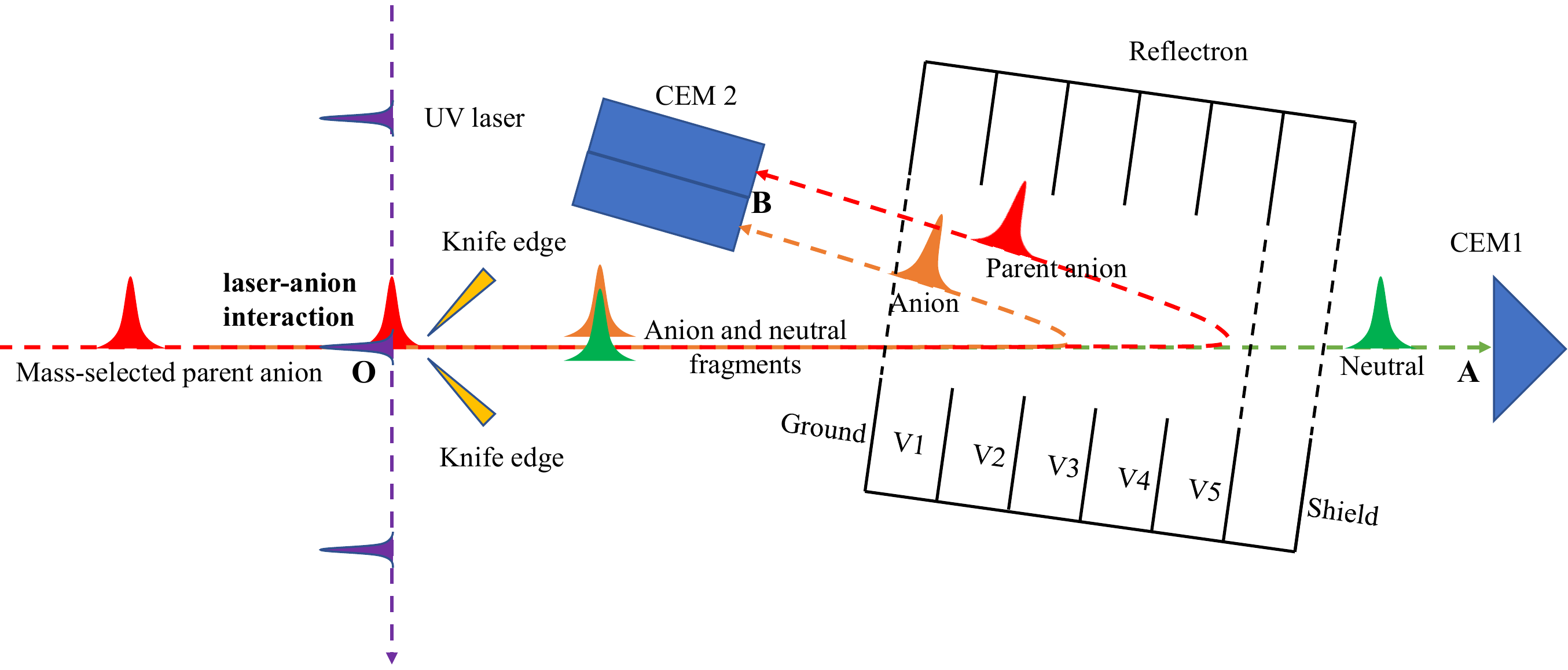}
\caption{Top view diagram of the detection chamber. The mass-selected parent ion-molecule cluster (red) is crossed by the UV beam (purple) at position O, marked as "laser-anion interaction". Anion fragments (golden) and the parent anion (red) are reversed by reflectron and hit detector CEM2 at position B. Neutrals(green) pass through the reflectron and hit detector CEM1 at position A. The parent anion has higher kinetic energy and hits upper detection area of CEM2. Anion fragments have lower energy and hit the lower detection area of CEM2. The dashed black lines of the reflectron indicates the high-transparency conductive meshes. }
\label{detection}
\end{figure*}

As shown in Fig.~\ref{detection}, the mass-selected ion-molecule cluster pulse is excited by an ultraviolet (UV) femtosecond laser pulse which propagates in a direction perpendicular to the ion beam. The pulses overlap at the interaction region (position O in Fig.~\ref{detection}). The detection chamber contains a copper pinhole (L in Fig.~\ref{overview}) for aligning the anion and laser beams and a pair of knife edges mounted on translation stages for measuring the transverse profile of the anion beam. The anion fragments produced after the laser excitation are mass-resolved using a reflectron mass spectrometer and captured using a CEM detector (CEM2). The neutral fragments pass through the reflectron and are captured by a CEM detector (CEM1) (see Fig.~\ref{detection}). The operating pressure of the detection chamber is $2\times10^{-7}$~Torr. The design details of each part here are discussed as follows.

The spatial overlap of the laser and anion beams is set using a 3 mm by 4.2 mm elliptical copper pinhole that is oriented at 45{\textdegree} relative to the direction of both beams. The laser beam size on the pinhole is adjusted to maximize the signal of neutrals on CEM1 while minimizing the background of low-energy photoelectrons generated by the UV light. A pair of movable knife edges with CEM1 (N in Fig.~\ref{overview}) is used to measure the cluster beam size in the transverse plane. The 12.7~mm-long knife edges, placed 9~mm downstream after pinhole, are mounted on two linear feedthroughs which are perpendicular to each other (see Fig.~\ref{detection}). The counts of the cluster beam passing through are recorded by CEM1 while the knife edges are scanned until fully blocking the cluster beam. An error function is used to fit the counts of the cluster beam as a function of the position of the knife edge, and the size of the cluster beam is calculated to be 2.90~mm$\times$3.28~mm (FWHM) in transverse plane.

A LPQR\cite{oh_tandem_2004} is employed to mass-resolve the anion fragments and the parent anion (see Fig.~\ref{detection}). To understand the mode of the LPQR's operation, we first examine how a LPQR refocuses anions with a small energy spread. Considering two anions with same mass, the anion with higher energy penetrates the reflectron deeper and thus spends more time inside the reflectron. After both anions come out of the LPQR, the anion with higher energy catches and overtakes the anions with lower energy. Thus, there is a focal position where anions of the same mass arrive at the same time.   

The main advantage of the LPQR compared to a linear reflectron\cite{alexander_recombination_1988}, is that the focal position for anions of a certain kinetic energy is approximately constant when the ratio between the quadratic and linear parts of the potentials applied to the LPQR electrode is a fixed value which depends on the kinetic energy of the anion, length of the LPQR, and the distance between the TOF mass spectrometer focus and the entrance of the LPQR. The detector CEM2 (O in Fig.~\ref{overview}) is positioned at the focus. To determine the focus position, we first calculate the TOF of the parent anion with energy U from the focus of the TOF mass spectrometer (O in Fig.~\ref{detection}) to a position after the LPQR. Here we follow the derivations given before\cite{oh_tandem_2004} and ignore the effects from the LPQR's small ($\sim$ 8$^\circ$) rotation relative to the anion beam direction. The potential inside the LPQR $V(x)$ has linear and quadratic components $V(x)={c_1}x+{c_2}x^2$, where $x$ is the distance from the entrance of the LPQR. The length of the reflectron is $d_r$ and we define $V_1={c_1}d_r$ and $V_2={c_2}{d_r}^2$. The maximum potential $V$ has the form (on the last electrode of the reflectron): $V=V_1+V_2$. Thus, the TOF for an anion with a kinetic energy of $U$, mass $m$ and charge q from the focus of TOF assembly to a certain position in front of the LPQR becomes\cite{oh_tandem_2004}
\begin{equation} \label{eq:1}
    T =\frac{d_1+d_2}{\sqrt{2U/m}}+\sqrt{\frac{2m}{q c_2}}\left[\frac{\pi}{2}-\sin^{-1}\left(\frac{1}{\sqrt{1+4c_2U/q{c_1}^2}}\right)\right],
\end{equation}
where $d_1$ is the distance from the TOF mass spectrometer's focus to the LPQR's entrance and $d_2$ is the distance from the LPQR's entrance to a certain position in front of the LPQR. The condition $\diff T$/$\diff U$ $= 0$ is used to find the focus equation: 
\begin{equation} \label{eq:2}
    \frac{d_1+d_2}{d_r} = \frac{4\langle{U}\rangle V_1}{q{V_1}^2+4\langle{U}\rangle V_2}.
\end{equation}
Here $\langle{U}\rangle$ is the average kinetic energy of the anions of mass m and charge q. 

Using the focus equation, we explain some of the differences between linear, quadratic, and linear-plus-quadratic reflectrons. After photoexcitation, the parent anion and the anion fragments have the same speed, and thus different kinetic energies that are proportional to their masses. The LPQR cannot focus the parent and fragment anions at the same position unless $d_1 +d_2 = 0$, which is automatically satisfied for the quadratic reflectron ($c_1 = 0$ and $V_1=0$). In this special case, the positions of the detector and the interaction volume overlap at the entrance of the reflectron, which is not practical in an experimental setup. In the linear reflectron ($c_2 = 0$), the focus equation changes to $d_2 = 4\langle{U}\rangle/{qE}-d_1,$ here $E = V_1/d_r,$ which is the electric field inside the linear reflectron. If the potential $V_1$ varies, the focus condition is lost for anions of average energy of $\langle{U}\rangle$ since $d_1$ is set. However, in the LPQR the focal position stays unchanged for a given anion while the potential applied to the reflectron electrode varies. In our LPQR setup, $d_1$, $d_2$ and $d_r$ are set to be 65~mm, 12~mm and 64~mm, respectively. For the parent anion with a kinetic energy of 2737~eV, the focus equation for the LPQR gives $V_2=-9\times10^{-5}{V_1}^2 + 0.83V_1.$ We make an approximation to get $V_2 = 0.83V_1$, $c_1 = 77c_2$ when $V_1$ is small. Thus, the potential inside the LPQR is $V(x) = 1.11\times 10^{-4} V (77x+x^2)$. If the potentials that are applied to electrodes of the LPQR follow this relation, the focus equation is valid, and the focal position remains unchanged. A homemade DC voltage divider is used to apply the prescribed voltages to the electrodes such that the focal position of the parent anion does not change if the potential V on the last electrode of the LPQR is varied.

The LPQR is mounted such that the angle between the reflectron and the incoming anion beam can be adjusted using a rotary feedthrough and a stepper motor. The angle is set to {8\textdegree} to accommodate the detector (CEM2) 12~mm in front of the reflectron. The LPQR consists of 6 stainless steel planar ring electrodes (Ground and V$_1$ to V$_5$ in Fig.~\ref{detection}) with 110~mm outer diameter, 50 mm inner diameter and 1.6~mm thickness, and with fixed spacing of 11~mm between the electrodes. The reflectron is enclosed by a cylindrical tube to confine the electrostatic field. High-transparency (79$\%$) conductive meshes cover the apertures of the first and last electrodes as well as the shielding tube. The length of the LPQR is 64 mm as mentioned previously. Under these operating conditions, the typical temporal length of the parent anion cluster is 100 ns (FWHM) on CEM2. We normally attain a fragment resolution of m/$\Delta$m$\approx$16, by considering the width of the detected ion peak corresponding to anion mass 127 (I$^-$) atomic mass units (amu) created by laser excitation of the CF$_3$I$\cdot$I$^-$ (323 amu) cluster.   

\subsection{Anion pulse-laser pulse synchronization and detectors}

\begin{figure}
\centering
\includegraphics[width=0.48\textwidth]{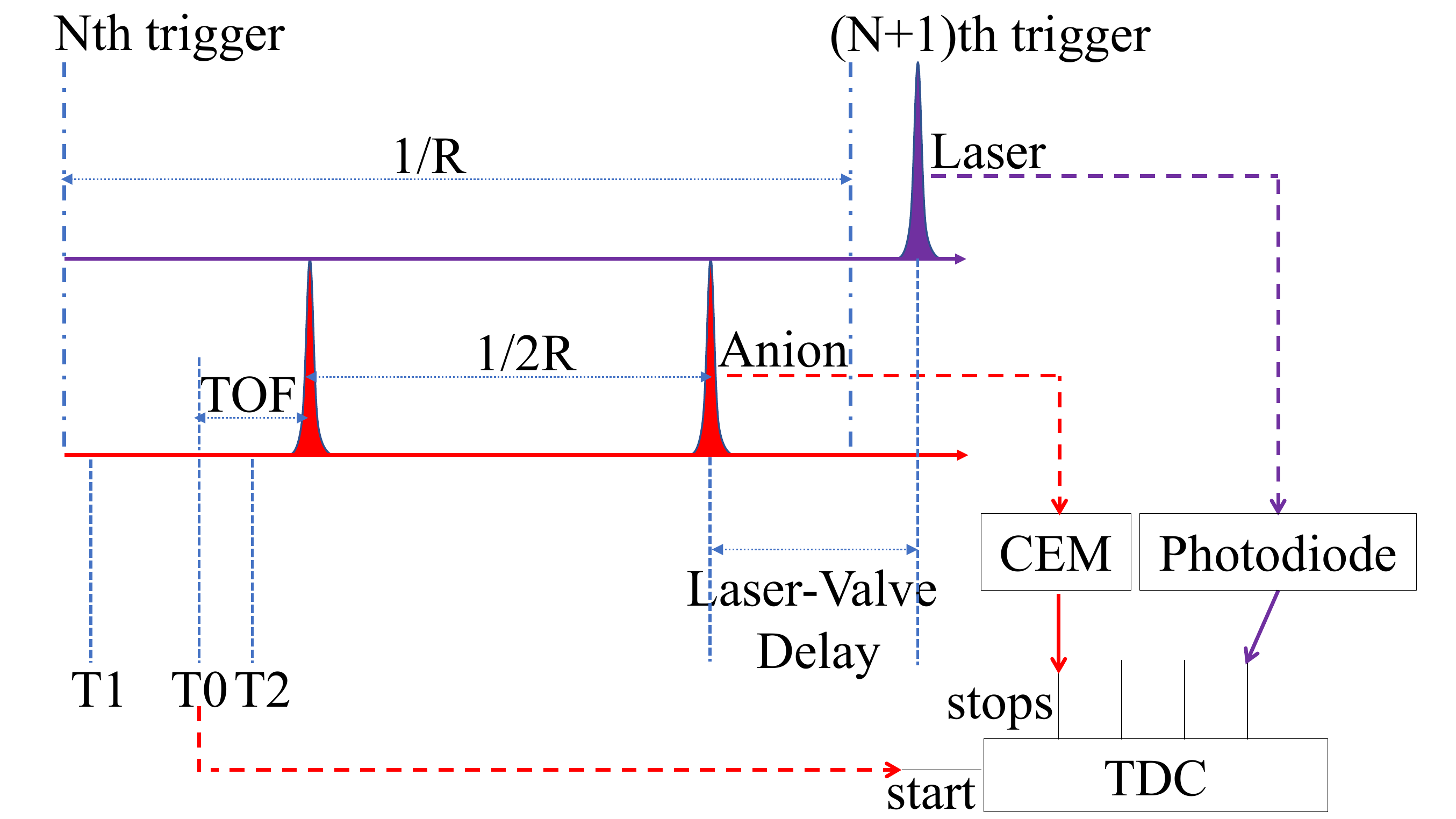} 
\caption{Overview of the synchronization system. T0 is the trigger for the TOF mass spectrometer. And it is also the start signal for the TimeTagger TDC and thus the origin of time axis for all TOF spectrum. T1 and T2 are triggers for Even-Lavie pulsed valve and mass filter respectively. Adjustable laser-valve delay is used to control the arrival time of the anion pulse relative to the laser pulse and overlap the anion pulse with the laser pulse. R is the repetition rate of the laser and set to be 100 in the experiments.}
\label{timing}
\end{figure}

A timing system is set up to synchronize the arrival of the anions and laser pulses at the interaction point. This requires precise timing of the nozzle opening and the HV pulses in the TOF mass spectrometer and mass filter. All time delays are set with respect to the laser trigger signal. Detectors CEM1 and CEM2 are used to monitor fragments after photoexcitation to optimize the temporal overlap. In this section, we will discuss the detectors and describe the timing system (see Fig.~\ref{timing}). 

Detectors CEM1 and CEM2 are used to record neutral and anion fragment signals, respectively. CEM1 (KBL 10RS/90-EDR, Dr.Sjuts Optotechnik GmbH) is a single CEM with a circular opening of 10 mm in diameter. It is used to detect neutral signals when the reflectron is active, or ion signals when the reflectron is inactive. The TOF spectrum from CEM1, shown in Fig.~\ref{tof_mass_filter}, is acquired when the reflectron is inactive. CEM2 is an array of two CEMs, each with a rectangular opening of size 5 mm$\times$15 mm (KBL 1505-EDR, Dr.Sjuts Optotechnik GmbH), with a total detection area of 5 mm$\times$30 mm. CEM2 is particularly useful for detecting fragment ions distributed over a large area at the detector. Small potentials -60 V and -15 V are applied to the inputs of CEM1 and CEM2 respectively to discriminate against a low-energy electron background generated from nearby surfaces by the UV laser and, to a lesser extent, collisions between ions and the wires comprising each mesh. Constant potentials of +2900 V and +2400 V are applied to the anodes of CEM1 and CEM2 respectively to amplify and extract electron pulses, which requires a capacitive coupling of signals to the pulse processing electronics. The extracted signals are amplified by a 4-channel preamplifier (SR240A, Stanford Research Systems) and discriminated by a 4-channel constant-fraction discriminator (21X4141,Lawrence Berkeley Laboratory). The resulting digitized signals, one from CEM1, and two from CEM2 (see Fig.~\ref{detection}), are sent to the first three stop channels of the TimeTagger (RoentDek TDC4HM-lr1) time-to-digital (TDC) converter . The TDC converter has a temporal resolution of 0.5 ns with no dead time. The temporal range, acquisition time and bin size of each TOF spectrum are adjustable and controlled by a computer. The computer also reads data from the TDC, displays it in TOF spectrum and stores the data on a disk which can be extracted for further analysis. The last stop channel of the TDC is used for recording the output signal from a fast photodiode (SM05PD2A, Thorlabs, Inc) that records the time of arrival of the UV laser pulse. 

The anion and laser pulses are synchronized using the timing system shown in Fig.~\ref{timing}. The trigger signal from the laser is used as the master and all delays are set with respect to this signal. The repetition rate of the master is doubled to 200 Hz by a delay/pulse generator (SRS DG535, Stanford Research Systems), and the output is used as the trigger T1 for the Even-Lavie valve generating anion pulses. Thus, the period (1/R) of the laser pulse is twice than that of the anion pulse (1/2R) in Fig.~\ref{timing}, where R is the repetition rate of the laser pulse. The time difference between T1 and Nth trigger in Fig.~\ref{timing} is the insertion delay of the delay/pulse generator. T0 is used as the trigger for the TOF mass spectrometer to extract and accelerate the ions, and as the start signal for the TDC. Thus, T0 is the temporal origin of the TOF spectrum shown previously in Fig.~\ref{tof_mass_filter}. T0 is delayed by 0.47 ms relative to T1, which is the transport time of the ions from the ionization region to the extraction region of the TOF mass spectrometer. T2 is delayed by 16.590 {\textmu}s relative to T0, and used as the trigger for the mass filter to select target anions. In the experiments, the UV laser pulse arrives at the optical window 2.67 {\textmu}s after the master (Nth trigger in Fig.~\ref{timing}) measured by the photodiode, much earlier than the anion pulse without any delays. Thus, we need to delay the arrival of the anion pulse to overlap it with next laser pulse after (N+1)th trigger. Laser-valve delay in Fig.~\ref{timing} can be adjusted to temporally overlap the anion pulse with laser pulse, and it is achieved by adding a delay between the Nth trigger and T1 trigger. This delay can be calculated as 1/R +2.67us-T0-TOF-1/2R. Varying the laser-valve delay changes the timing of the trigger for the Even-Lavie pulsed valve, and thus controls the arrival time of anion pulse relative to the laser pulse.   

The detector CEM2 is placed at position O (see Fig.~\ref{detection}) to measure the time of flight for the parent anion before the photoexcitation experiment. The arrival time of the parent anion from the detector CEM2 and the arrival time of the laser pulse from the photodiode are processed by TDC and displayed as TOF spectrum. The optimized laser-valve delay is found when signals from the UV laser pulse and the parent anion pulse are overlapping on the TOF spectrum. The neutral signals from photofragments are very sensitive to the change of laser-valve delay, and thus they act as an effective indicator for the overlap between laser and anion pulses.

\section{Data analysis and results}

We test the performance of the apparatus by measuring the photofragments after photoexcitation of the ion-molecule cluster CF$_3$I$\cdot$I$^-$. We excite the cluster CF$_3$I$\cdot$I$^-$ using the UV laser pulses with a wavelength of 266 nm. The intracluster charge transfer process after the photoexcitation is investigated by detecting all the possible photofragments: anion fragments and neutrals. The UV laser pulse with energy of 110 {\textmu}J is generated by the 3rd harmonic conversion of a Ti:sapphire laser with central wavelength of 800 nm, temporal length of 50 fs and 100 Hz repetition rate. Photofragment data is acquired at different laser-valve delays with a step of 40 ns which corresponds to half of the duration of the CF$_3$I$\cdot$I$^-$ pulses at position O (shown in Fig.~\ref{detection}). Alternating parent cluster pulses are photoexcited by the UV laser pulse, and thus a real-time CF$_3$I$\cdot$I$^-$ cluster background (parent anion without laser crossing) is recorded. The laser background is recorded separately for the same acquisition time of 1000 s with the pulsed valve switched off. The parent anion and laser backgrounds are subtracted from the raw fragment signal. The fragment signals are normalized to the parent anion signals for each laser-valve delay. The anion signal is captured by CEM2 at position B, while the neutral fragment signal is captured by CEM1 at position A (position A and B shown in Fig.~\ref{detection}). 

In the next two sections, we discuss how to identify the photofragments with the help from simulations of the anion trajectories, and we compare our results to previous results and discuss the dynamics of photoexcitation of anion cluster CF$_3$I$\cdot$I$^-$.   

\subsection{Photofragments of CF$_3$I$\cdot$I$^-$: anion and neutral fragments} 

The well-known relation that mass is a quadratic function of flight time is normally used to fit the experimental data to search and identify the fragment peak in the TOF mass-resolving experiment. However, the relation is not valid in the LPQR measurement due to the quadratic terms of the potentials applied to the LPQR (see Eq.~\ref{eq:1}). Thus, we search and identify the anion fragment signals by comparing the experimental data to SIMION\cite{dahl2000simion} calculations. 

\begin{table} 
\caption{\label{tab:table} TOF (in {\textmu}s) data of parent anions, anion fragments and neutrals at different position A and B starting from position O in the experiment and SIMION simulation.}
\begin{ruledtabular}
\begin{tabular}{cccc}
\mbox{Molecule}&\mbox{T@A(expt.)}&\mbox{T@B(expt.)}&\mbox{T@B(sim.)}\\
\hline
\mbox{CF$_3$I$\cdot$I$^-$}&\mbox{4.350}&\mbox{5.815}&\mbox{5.90}\\
I$^-$(source)& 2.655 & 3.602 & 3.70 \\
I$^-$(fragment)& NA\footnotemark[1] & [3.6  4.0] & 3.82 \\
CF$_3$I$^-$& NA\footnotemark[1] & [4.5  4.9] & 4.65 \\
Neutral&[4.1 4.5]&NA\footnotemark[2]&NA\footnotemark[2] \\
\end{tabular}
\end{ruledtabular}
\footnotetext[1]{There is no anion fragment signal on CEM1 at position A when the reflectron is on.}
\footnotetext[2]{Neutrals are not reflected by the reflectron and will not hit CEM2 at position B.}
\end{table}

Table~\ref{tab:table} lists the TOFs from position O to A and B for CF$_3$I$\cdot$I$^-$ and I$^-$ generated in source chamber, possible anion fragments I$^-$ and CF$_3$I$^-$, and neutrals produced by photoexcitation. TOFs of parent anions CF$_3$I$\cdot$I$^-$ and I$^-$ (generated in source chamber) at position O can be measured using CEM2 with the mass filter off, at position A using CEM1 with the reflectron and mass filter both off and at position B using CEM2 with the mass filter on and reflectron off. The distance between positions O and A is 175.55 mm, and the speed of parent anion CF$_3$I$\cdot$I$^-$ is calculated to be 40.4 mm/{\textmu}s. Possible anion fragments I$^-$ and CF$_3$I$^-$ have the same speed and size as parent anion at position O where photoinduced fragmentation occurs. Given the initial conditions measured experimentally, the trajectories for possible anion fragments I$^-$ and CF$_3$I$^-$ from position O to B with the reflectron on can be simulated accurately using SIMION. The extracted TOFs from the simulation for possible anion fragments I$^-$ and CF$_3$I$^-$ are 3.82 {\textmu}s and 4.65 {\textmu}s respectively. The SIMION simulations accurately describe the data, with small deviations that are smaller than 100ns, the duration of the CF$_3$I$\cdot$I$^-$ pulses on CEM2. 

\begin{figure}[!htb]
\centering
\includegraphics[width=0.48\textwidth]{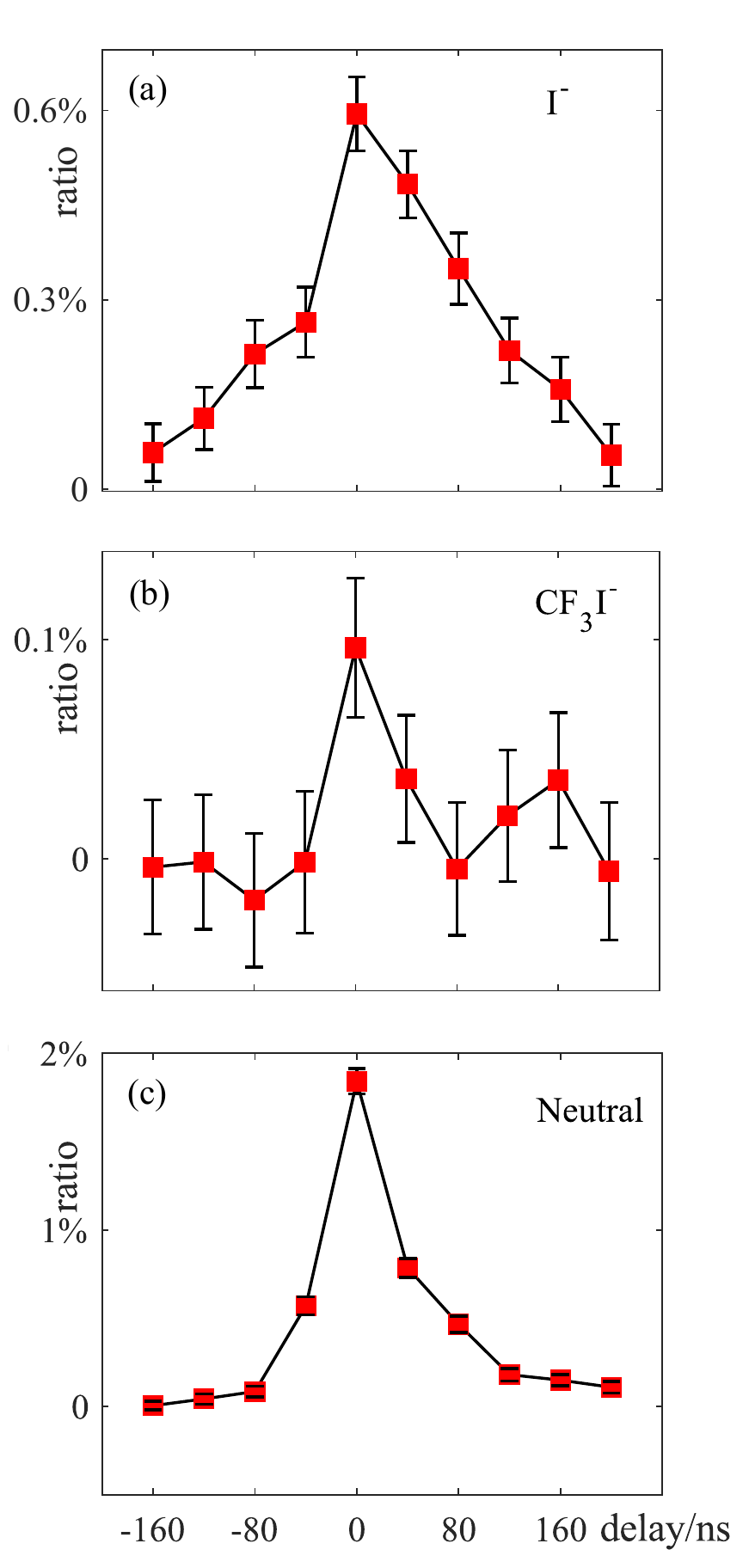}
\caption{(a), (b) and (c) show the ratio at different delays of photofragments: I$^-$, CF$_3$I$^-$ and neutrals respectively. Ratio here is defined as fragment signal divided by parent anion CF$_3$I$\cdot$I$^-$ signal. Photofragments shown here are generated after photoexcitation of parent ion-molecule cluster CF$_3$I$\cdot$I$^-$ crossed by 266 nm UV pulse. Delay step in the data is 40 ns and optimized laser-valve delay is 4ms531{\textmu}s110ns which is set to be 0 as reference timing in figure (a), (b) and (c).}
\label{data}
\end{figure}

We used the simulated flight time to search and identify the possible anion fragments generated in the photofragmentation of CF$_3$I$\cdot$I$^-$. Based on the simulated TOF 3.82 {\textmu}s from position O to position B, TOF range of [3.6 {\textmu}s 4.0 {\textmu}s] (TOF of parent anion at position O included) is thus chosen to be the integration range for the fragment I$^-$ recorded by CEM2. The I$^-$ signal is normalized to the counts of the parent anion CF$_3$I$\cdot$I$^-$ to account for variations in the incoming anion beam current, and shown in Fig.~\ref{data}(a). The I$^-$ signal reaches its peak at laser-valve delay of 4ms531{\textmu}s110~ns. For the CF$_3$I$^-$ fragment, counts are integrated in the interval [4.5 {\textmu}s, 4.9 {\textmu}s]. The normalized signal is shown in Fig.~\ref{data}(b) which reaches maximum at the same laser-valve delay as the I$^-$ signal. The accompanying neutral fragments have the same arrival time at position A as the parent anion cluster CF$_3$I$\cdot$I$^-$ measured to be 4.350 {\textmu}s. The integration interval for the neutral fragments is [4.1 {\textmu}s, 4.5 {\textmu}s]. The corresponding normalized signal is shown in Fig.~\ref{data}(c). The maximum signal for neutral fragments corresponds to the same delay as that for anion fragments I$^-$ and CF$_3$I$^-$, which verifies that both neutrals and anion fragment I$^-$ and CF$_3$I$^-$ are generated during the photoexcitation of parent anion CF$_3$I$\cdot$I$^-$. 

\subsection{Comparison with calculations and discussion}

We have carried out quantum chemical electronic structure calculations with the General Atomic and Molecular Electronic Structure System (GAMESS) software package\cite{barca2020recent} to aid the interpretation of our experimental results. Second-order Møller-Plesset perturbation theory (MP2)\cite{moller1934note,binkley1975moller} methods\cite{ishimura2006new,ishimura2007new,aikens2006scalable,fletcher2002gradient} are used for molecules in their electronic ground states (either closed shell or open shell wave functions). The completely renormalized equation-of-motion coupled-cluster singles, doubles and noniterative triples (CR-EOM-CCSD(T)) method\cite{piecuch2002efficient,piecuch2009left,kowalski2004new} is used to calculate excited state wave functions and energies. The Sapporo non-relativistic basis sets\cite{tatewaki1996contracted,tatewaki1997contracted,koga1999contracted,koga2002contracted} are used in the MP2 and CR-EOM-CCSD(T) calculations. Here, SPK-ADZP denotes the augmented polarizable double zeta set, SPK-ATZP denotes the augmented polarizable triple zeta set, and SPK-AQZP denotes the augmented polarizable quadruple zeta set.

The photoexcitation of CF$_3$I$\cdot$I$^-$ was recently studied by photoelectron spectroscopy and density functional theory\cite{mensa-bonsu_photoelectron_2019}. The two direct photodetachment thresholds of CF$_3$I$\cdot$I$^-$ are 4.0 eV and 4.9 eV\cite{mensa-bonsu_photoelectron_2019}, the difference of which corresponds to the spin-obit splitting of the iodine atom. The difference between the lowest vertical detachment energy (VDE) of the cluster and the electron affinity of iodide ion is known as the stabilization energy\cite{ayala2004study}. Considering that the electron affinity of iodine is 3.059 eV\cite{pelaez_pulsed_2009}, the stabilization energy of the CF$_3$I$\cdot$I$^-$ anion cluster is around 0.9 eV which is much higher than many other anion clusters: 0.36 eV for I$^-$$\cdot$CH$_3$I\cite{mabbs_photoelectron_2005}, 0.46 eV for I$^-$$\cdot$H$_2$O\cite{mabbs_photoelectron_2005}, 0.47 eV for I$^-$$\cdot$CH$_3$CN\cite{mabbs_photoelectron_2005}, 0.1 eV for I$^-$$\cdot$CO\cite{doi:10.1021/jp300471x} and 0.172 eV for I$^-$$\cdot$CO$_2$\cite{doi:10.1063/1.468576}. The relatively high stabilization energy of CF$_3$I$\cdot$I$^-$ implies that there is a strong interaction between the molecule and I$^-$ in the cluster. The electronic structure calculation at MP2/SPK-ATZP level of theory for CF$_3$I$\cdot$I$^-$ shows that in the electronic ground state of CF$_3$I$\cdot$I$^-$, the I$^-$ ion directly binds to the I atom, which has a $\sigma$-hole thus positively charged\cite{clark2007halogen}. In CF$_3$I$\cdot$I$^-$ the I-I distance is calculated to be 3.314~\AA, which is similar to the calculated 3.230~\AA~in {I$_2$}$^-$. These results are similar to those reported earlier\cite{mensa-bonsu_photoelectron_2019}. The binding energy of I$^-$ to CF$_3$I calculated at the MP2/SPK-AQZP//MP2/SPK-ATZP level of theory is 0.742~eV. 



Johnson {\it et al.} \cite{cyr_observation_1992,cyr_photoelectron_1993,cyr_charge_1994,dessent_vibrational_1996,doi:10.1021/j100006a002} conducted a detailed investigation of anion cluster I$^-$$\cdot$CH$_3$I using photoelectron spectroscopy, photofragmentation action spectroscopy and translational spectroscopy which showed that a charge-transfer excited state exists just below the VDE. Photofragmentation accompanies excitation to the charge-transfer excited state, and the photon energy-dependent photofragment (I$^-$ and I$_2^-$) yields peak at a photon energy of 3.42 eV, which is 0.014 eV below the direct detachment threshold for I$^-$$\cdot$CH$_3$I\cite{cyr_charge_1994}. Later Scarton {\it et al.} conducted a photoexcitation experiment using 260 nm (4.77 eV) and 318 nm (3.90 eV) laser pulses to excite the cluster CF$_3$I$\cdot$I$^-$ below each of the two direct photodetachment thresholds, and the fragment CF$_3$I$^-$ was detected under both wavelengths\cite{scarton_study_nodate}, which is consistent with the present results.

In the present experiments, we use 266 nm (4.66 eV) femtosecond UV pulses to initiate the photofragmentation of CF$_3$I$\cdot$I$^-$, and directly detect the photofragments I$^-$, CF$_3$I$^-$, and total yield of neutrals, as shown in Fig.~\ref{data}. The photon energy we used here is well above the lower direct detachment energy 4.0 eV and just below the higher direct detachment energy 4.9 eV of CF$_3$I$\cdot$I$^-$. The dominant anion product is I$^-$, which we will discuss after considering the second most abundant anion product, CF$_3$I$^-$. 


The detection of the anion fragment CF$_3$I$^-$ and neutrals, resulted from UV photoexcitation of CF$_3$I$\cdot$I$^-$, implies that electronic excited states of CF$_3$I$\cdot$I$^-$ are accessed in the present experiment with a fixed photon energy of 4.66 eV. Excited state electronic structure calculation performed at CR-EOM-CCSD(T)/SPK-ADZP level of theory suggests that there are several charge-transfer singlet excited states in (CF$_3$I$\cdots$I)$^-$: an electron in one of the two p orbitals of the anion I$^-$ can transfer to the s and p orbitals of the I atom in CF$_3$I upon photoexcitation. Therefore, the molecular anion in these excited states can be denoted as CF$_3$I$^-$$\cdot$I. The excitation energies are calculated to be 4.18 (degeneracy=2), 4.38, 4.73 (degeneracy=2), 4.85~eV. The higher excited states are short-lived states so only the first excited state (4.18~eV, the degenerate S1 state here) are chemically meaningful when fragmentation is considered. CR-EOM-CCSD(T)/SPK-ADZP single point energy calculations performed at various I-I distances (from 3.314 Å to 6.0 Å) for CF$_3$I$^-$$\cdot$I clearly show that, on the S1 excited state potential energy surface, this molecular anion is not stable and will dissociate into CF$_3$I$^-$ and I. The energy lowering in this dissociation process (from 3.314 Å to 6.0 Å) is calculated to be 0.812 eV . 
After CF$_3$I$^-$$\cdot$I dissociates into CF$_3$I$^-$ and I, the CF$_3$I$^-$ anion can further dissociate into CF$_3$ and I$^-$, with a relatively low energy barrier of 0.324 eV (as calculated at the MP2/SPK-AQZP//MP2/SPK-ATZP level of theory). This barrier can be easily overcome when CF$_3$I$^-$ is produced from the fragmentation of CF$_3$I$^-$$\cdot$I, which releases at least 0.812~eV energy in forms of translational kinetic energy of both CF$_3$I$^-$ and I, rotational kinetic energy of CF$_3$I$^-$, and internal vibrational excitation in CF$_3$I$^-$. While a detailed theoretical analysis on the fragmentation energy distribution cannot be easily performed, there is a high probability that the internal vibrational energy attained by CF$_3$I$^-$ exceeds 0.324 eV. Therefore, the molecular anion is metastable with respect to dissociation by C-I break, producing a detectable I$^-$ ion in the present experiment, as shown in Fig. 6(a). There appears to be a lower probability that CF$_3$I$^-$ attains an internal vibration energy less than 0.324 eV, and cannot dissociate, remaining as detectable CF3I- anions, as shown in Fig. 6(b). So not all CF$_3$I$^-$ anions generated from photofragmentation of CF$_3$I$^-$$\cdot$I can dissociate into CF$_3$ and I$^-$. We note dissociative electron attachment (DEA) experiments\cite{marienfeld_high_2006,heni_dissociative_1986} on CF$_3$I, have reported no detectable CF$_3$I$^-$, i.e., all CF$_3$I$^-$ anions dissociate into CF$_3$ and I$^-$ following resonant attachment of free electrons with energies as low as 0.5 meV. This is due to the energy of the ground state of CF$_3$I being above both the CF$_3$ + I$^-$ dissociation limit of the CF$_3$I$^-$ anion, and any barriers to dissociation. 

Therefore, the atomic I$^-$ anions detected in the current work are from CF$_3$I$^-$ fragmentation. They are not from the direct dissociation of the 4.66 eV photon excited parent CF$_3$I$^-$$\cdot$I anions, in accordance with the photoelectron spectroscopic experiments on CF$_3$I$\cdot$I$^-$\cite{mensa-bonsu_photoelectron_2019}. In addition, they are not from the fragmentation of the unexcited parent CF$_3$I$\cdot$I$^-$ anions. As discussed above, the binding energy between CF$_3$I and I$^-$ in CF$_3$I$\cdot$I$^-$ is calculated to be 0.742 eV, which is high enough to suppress the dissociation.

\section{Summary}

In summary, we describe the design and implementation of an apparatus to investigate the charge transfer process and the resulting fragments after photoexcitation of ion-molecule clusters. The apparatus includes the generation of the ion-molecule clusters in a source chamber, separation and selection of the parent ion-molecule cluster using a TOF mass spectrometer and detection of photofragments after laser excitation of ion-molecule clusters. In our apparatus, a mass filter is employed to select a specific parent anion cluster for laser excitation. In addition, the LPQR is used to mass-resolve the anion photofragments. The performance of the apparatus is shown in experiments of the ion-molecule cluster CF$_3$I$\cdot$I$^-$ photoexcited by UV laser pulses with a wavelength of 266 nm UV. Anion fragments I$^-$ and CF$_3$I$^-$ are detected along with neutral fragments. The detection of CF$_3$I$^-$, supported by quantum chemical electronic structure calculations, suggests that charge transfer excited states are accessed in CF$_3$I$\cdot$I$^-$ cluster by 4.66~eV photons. CF$_3$I$^-$$\cdot$I in the charge transfer excited state is not stable and dissociates to produce CF$_3$I$^-$ and I. The energy released from this dissociation process can be partially attained by CF$_3$I$^-$ as internal vibration energy, enabling a majority of the CF$_3$I$^-$ anions to further dissociate into CF$_3$ and I$^-$.



\section{Acknowledgment}

We would like to thank Yanwei Xiong for his help with laser optics and Prof. Ilya Fabrikant for helpful discussions. Quantum chemical calculations were performed with resources at the University of Nebraska Holland Computing Center. This work was supported by the US Department of Energy (DOE), Office of Science (Sc), Division of Chemical Sciences Geosciences and Biosciences (CSGB) of the Office of Basic Energy Sciences (BES) under award no. DE-SC0019482. Work performed at Lawrence Berkeley National Laboratory was supported by the US DOE Sc BES CSGB under Contract DE-AC02-05CH11231. 


%

%


\bibliography{Instrumentalpaper.bib}

\end{document}